\documentclass[twocolumn,amsmath,amssymb,aps]{revtex4-1}

\usepackage{graphicx}
\usepackage{dcolumn}
\usepackage{bm}
\usepackage{amssymb}
\usepackage[T1]{fontenc}
\usepackage{currvita}
\usepackage{csquotes}

\begin{document}


%
%
\preprint{APS/123-QED}

\title{Reply to the Comments on the $^{12}$C+$^{12}$C fusion S$^*$-factor}



\author{A. Tumino$^{1,2,*}$\thanks{tumino@lns.infn.it}}\author{C. Spitaleri$^{2,3}$}\author{M. La Cognata$^{2}$}\author{S. Cherubini$^{2,3}$}\author{G.L. Guardo$^{2}$}\author{M. Gulino$^{1,2}$}\author{S. Hayakawa$^{2,4}$}\author{I. Indelicato$^{2}$}\author{L. Lamia$^{2,3}$}\author{H. Petrascu$^{5}$}\author{R.G. Pizzone$^{2}$}\author{S.M.R. Puglia$^{2}$}\author{G.G. Rapisarda$^{2}$}\author{S. Romano$^{2,3}$}\author{M.L. Sergi$^{2}$}\author{R. Spart\'a$^{2}$}\author{L. Trache$^{5}$}

\affiliation{$^1$Facolt\`a di Ingegneria e Architettura, Universit\`a degli Studi di Enna {\enquote{Kore}}, Enna, Italy}
\affiliation{$^2$INFN, Laboratori Nazionali del Sud, Catania, Italy\\ $^3$Dipartimento di Fisica e Astronomia, Universit\`a degli Studi di Catania, Catania, Italy}
\affiliation{$^4$Center for Nuclear Studies, The University of Tokyo, Tokyo, Japan}
\affiliation{$^5$Horia Hulubei National Institute for R$\&$D in Physics and Nuclear Engineering, Bucharest-Maguerele, Romania}
\affiliation{$^*$email: tumino@lns.infn.it\\}


\begin{abstract}
The goal of this reply is to draw attention of the readers that the major problems rose in the short Comment authored by A.M. Mukhamedzhanov, X. Tang and D.Y. Pang (https://arxiv.org/pdf/1806.05921.pdf) are totally groundless.
 \end{abstract}
 

\maketitle

The authors of the Comment posted under ArXiv \cite{Akram} have highlighted some points of our work published in Nature \cite{Tumino2018} that they consider weak or wrong and according to which they conclude that the result of our work is incorrect. The short Comment \cite{Akram} was followed by a more detailed discussion, authored by A.M.M. alone \cite{Akram1}. We will consider that paper as well in this reply. \\ We are surprised by the posted Comment also because some of the issues were discussed by some of us with A.M.M. and agreed. We think that the Comment was written in a hasty manner and we can see this from some errors which otherwise have no real explanation. In fact, they refer to this application of the Trojan Horse Method published in Nature as the first one. It is not the first application at all since the method has been applied several times since the early nineties and tested to check experimentally if it is reliable (\cite{Spitaleri2011,Tumino2013,Tribble2014,Spitaleri1999,Lacognata2005,Lacognata2011,Lamia13,Tumino14,Pizzone17,Spitaleri17,Dagata2018} and references therein). The authors of the Comment are perfectly aware of that since two of them have been coauthors of several of our papers. A.M.M. has authored more than $50$ papers with THM measurements (dating as early as 2005) and X.T. has authored $7$ papers (dating as early as 2013). \\ 

1) In our analysis, we did not take the d-$^{24}$Mg Coulomb interaction into account because, as demonstrated in the pioneering work reported in \cite{Dolinsky}, its effects can be safely neglected.
 In this paper and in many others authored by A.M.M., many of which are in collaboration with our group, the authors demonstrated that the plane-wave approximation provides the same energy dependence of the three-body cross section as the one obtained using the distorted-wave approximation, but in a much simpler way (yet, significantly departing from the absolute value). This fact is cited, for instance, in \cite{Tribble2014} (A.M.M. taking care of the theoretical section). Corrections were tested in several papers \cite{Pizzone2009,Guardo2017} with the help of A.M.M. himself, proving negligible in our phase space region. Also the pole invariance of the two-body cross section was investigated in different cases \cite{Tumino2006,Pizzone2011,Pizzone2013}.  It was demonstrated that, within experimental errors, leaving a neutron or a charged particle as spectator to the two body reaction does not affect its energy trend of the two-body cross section. \\
 Our cross section is the result of a high precision experimental work, where the appropriate tests have been performed. In the phase space region populated in our experiment, the $^{14}$N beam energy of 30 MeV corresponds to a quite high momentum transfer q$_t$=500 MeV/c with q$_t$ defined by the Galilean invariant equation reported in \cite{Tribble2014}, giving an associate de Broglie wavelength of 0.4 fm quite smaller than the $^{12}$C+d radius of about 3 fm \cite{Belyaeva01}. This substantiates the peripheral nature of the $^{14}$N+$^{12}$C interaction and the validity of the Impulse Approximation. Moreover, as thoroughly described in the Methods Section of \cite{Tumino2018}, under paragraph $``$Deuteron Momentum Distribution$"$, the agreement within experimental errors between the peculiar shape of the experimental deuteron momentum distribution and the theoretical one, clearly indicates that in the phase space region spanned in our experiment the plane-wave approximation can be relied on because no distortions are needed to describe our transfer process in the momentum window spanned in our work. This is consistent with the value of the Sommerfeld parameter for the d+$^{24}$Mg system, that turns out to be not larger than 1.03, similar to the values obtained in other THM works and in particular in our recent paper on $^6$Li+$^{19}$F at 6 MeV published in \cite{Dagata2018}. After discussion, A.M.M. agreed with us that the d+$^{23}$Na Coulomb interaction was not crucial. Thus, the general criticism raised by the authors about the need of a general theory does not apply to the present case. \\
However, the theory developed by A.M.M. was not validated before being applied to the $^{12}$C+$^{12}$C data \cite{Akram1}. The theory appears to fail in reproducing a fundamental aspect that characterizes the transfer process (regardless of its quasi-free nature) and that was observed in all previous experimental works some of them in similar kinematic conditions as our work (such as those published in \cite{Belyaeva01,Belyaeva02,Zurmuhle94}): for transfer process the deuteron angular distributions are peaked at forward angles. In the region at backward angles, the reaction cross section is typically dominated by the compound nucleus mechanism with little to no contribution from transfer. One of our referees was an expert of reaction mechanisms and he pretended from us quite a lot of supplemental analysis and material to show him/her unambiguous proof that selected data corresponds to direct $^{12}$C transfer. This is exactly the opposite of what A.M.M. states in his paper \cite{Akram1} and containing the basis of what he claims to be a general theory that uses the distorted-wave-born approximation (DWBA). In that paper A.M.M. states that the deuteron angular distribution from DWBA is peaked at backward angles, completely opposite to the Plane Wave Approach that he criticizes but which gives the forward peak according to experiment. Thus, this is sufficient to consider any application to experimental data, and in particular to $^{12}$C+$^{12}$C ones, to be meaningless.  A.M.M. is forgetting that our deuteron cannot go to backward angles being a projectile-like particle.
Maybe, this is why A.M.M. results drop too steep at low energy, while seem to diverge at higher energy, totally disagreeing with available direct data. The agreement with available direct data is a must for a theory to be reliable and this is not the case. The trend is shown in Figs.~1 and 2 for the $^{20}$Ne+$\alpha_1$ and $^{23}$Na+p$_1$  channels, respectively. 
In each figure, the solid black line plus shading is the THM S(E)$^*$ factor and related uncertainty published in \cite{Tumino2018}, colored symbols are direct data (red filled circles \cite{Spillane07}, purple filled squares \cite{Mazarakis73}, blue filled diamonds \cite{High77},  blue filled stars \cite{Kettner80} and green filled triangles \cite{Barron06}) and the solid blue line represents the theoretical calculation by A.M.M. \cite{Akram1}. The solid blue line has been extended at higher energy for comparison with direct data
since no comparison with direct data above 2.6 MeV is shown in \cite{Akram1}.
The procedure to extend the blue solid line at higher energy is straightforward: taking advantage of the agreement in the overlapping region between our THM S(E)$^*$ factor and the one derived from direct data, we consider the direct data as the extension at higher energy of our behaviour. Dividing the blue line by the black one, we determine the correction factor to be multiplied by the direct S(E)$^*$ factor above 2.7 MeV. The result is a diverging behaviour that does not reproduce the experimental one.\\
As for the rise of our S(E)$^*$-factors at low energies, it is obviously due to the resonant behaviour, while the baseline agrees within the experimental errors with available extrapolations neglecting the existence of low-lying resonances. Moreover, deviations from the reference rate \cite{CF88} at the temperature T$_9$=0.5 GK, is up to a factor of 30 and not 500 as claimed in \cite{Akram1}.\\
Finally, we do not see any shift in the resonance energy from the data. A.M.M. clearly states that the final-state three-body Coulomb interaction can shift the resonance energies.
However, the absence of any shift in the resonance energy means that the final-state three-body Coulomb interaction is not influential.\\
\begin{figure}[htp]
\includegraphics[width=1.01\columnwidth]{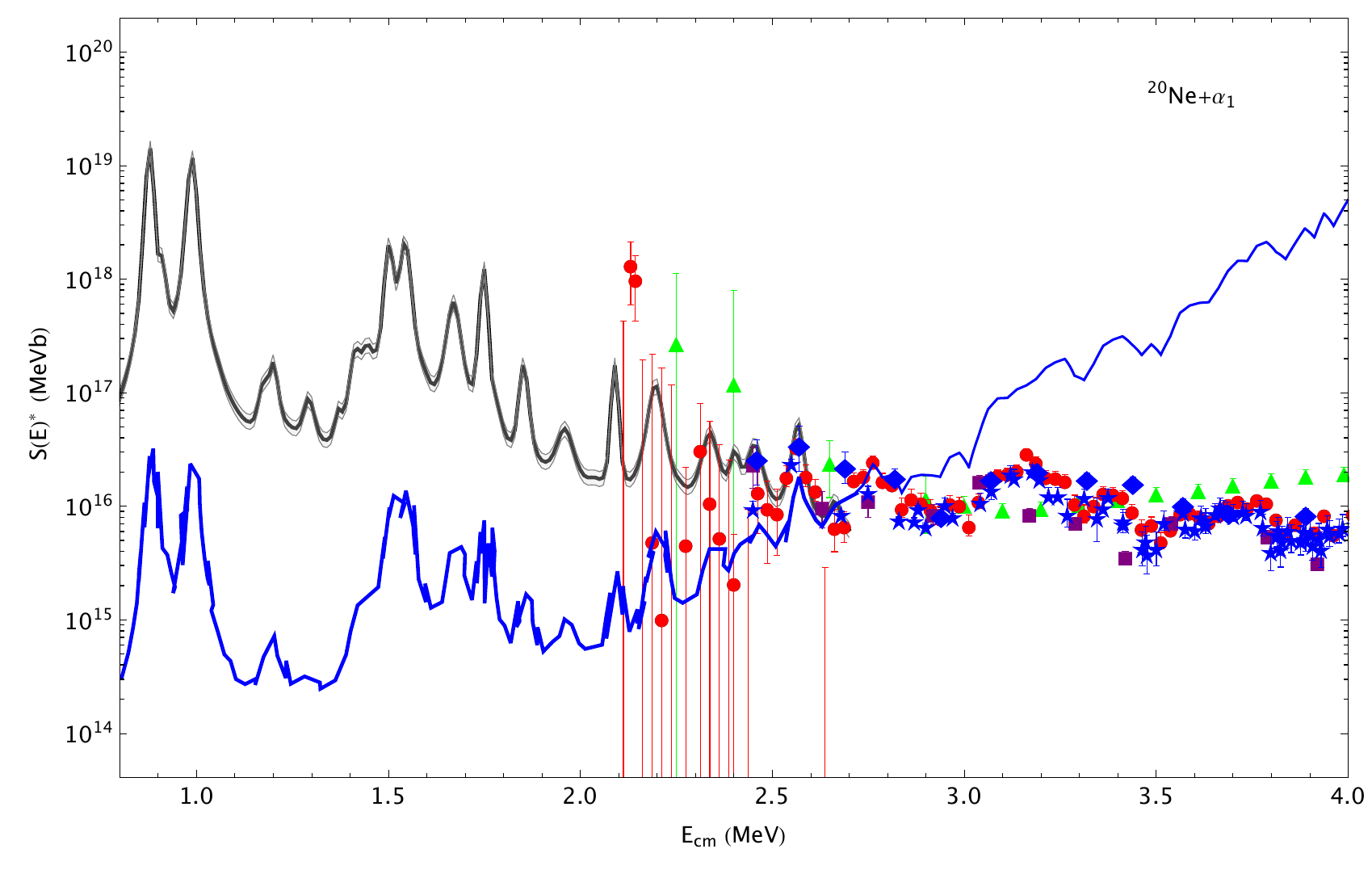}
\caption{Comparison between experimental and theoretical S(E)$^*$-factor for the $^{20}$Ne+$\alpha_{1}$ channel. The solid black line plus shading represent the THM S(E)$^*$ factor and related uncertainty published in \cite{Tumino2018}, coloured symbols are direct data (red filled circles \cite{Spillane07}, purple filled squares \cite{Mazarakis73}, blue filled diamonds \cite{High77},  blue filled stars \cite{Kettner80} and green filled triangles \cite{Barron06}) and the solid blue line represents the theoretical calculation by A.M.M. \cite{Akram1}. } \label{alfa3}
\end{figure}
\begin{figure}[htp]
\includegraphics[width=1.01\columnwidth]{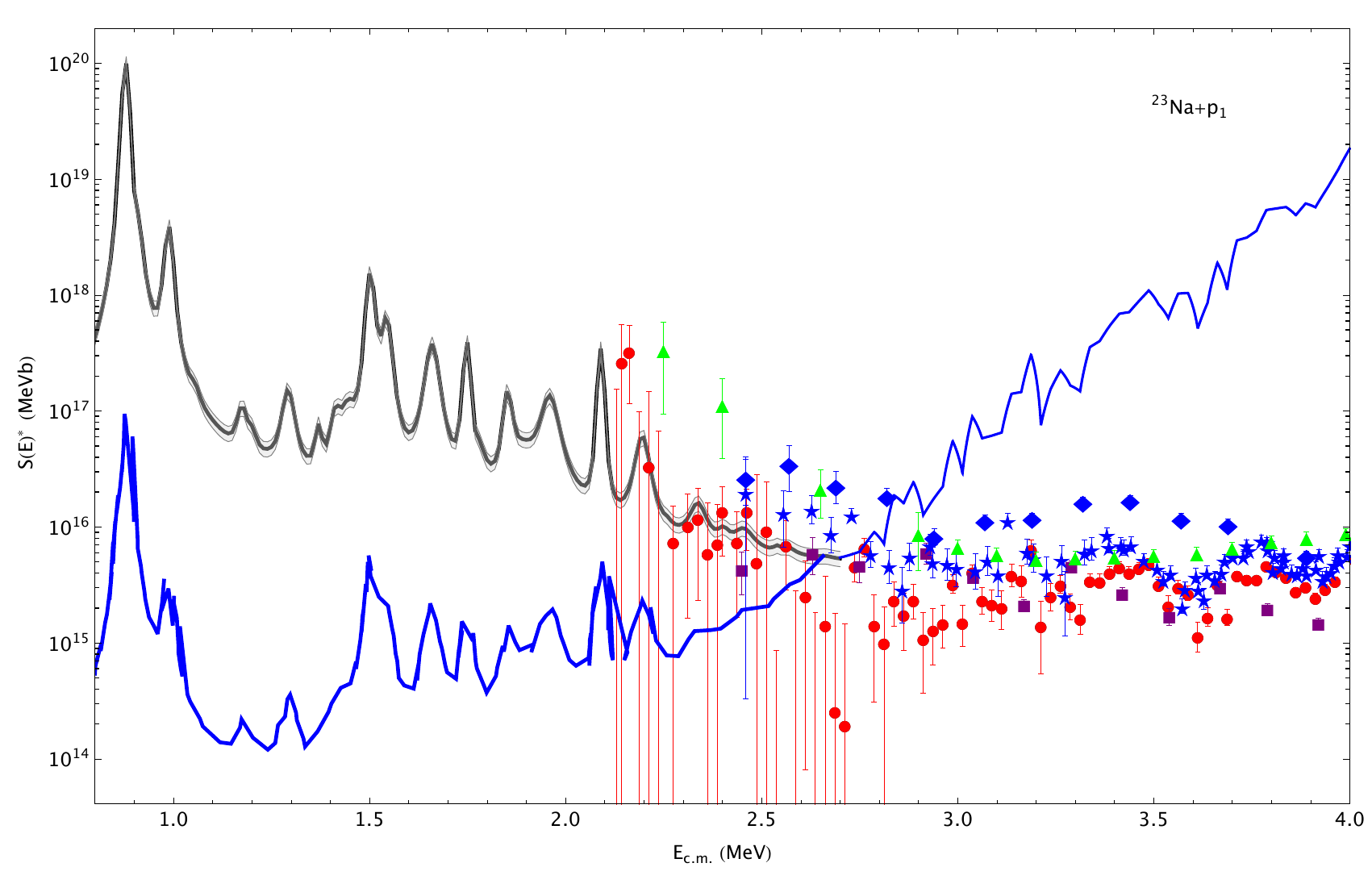}
\caption{Comparison between experimental and theoretical S(E)$^*$-factor for the $^{23}$Na+$p_{1}$ channel. The solid black line plus shading represent the THM S(E)$^*$ factor and related uncertainty published in \cite{Tumino2018}, coloured symbols are direct data (red filled circles \cite{Spillane07}, purple filled squares \cite{Mazarakis73}, blue filled diamonds \cite{High77},  blue filled stars \cite{Kettner80} and green filled triangles \cite{Barron06}) and the solid blue line represents the theoretical calculation by A.M.M. \cite{Akram1}.} \label{eij}
\end{figure}

2) Normalization to direct data was done using data sets available from literature in the region E$_{cm}$= 2.5-2.63 MeV. We have chosen this region since direct data are the most accurate among those available throughout the full overlapping range with THM data. Introducing different data sets has the effect of reducing the normalization error, leaving the most accurate ones to dominate in the procedure. This was also the advice of one of the referees. 
Now, concerning the so disputed data set of \cite{Mazarakis73}, they are not proved to be wrong, but there are only conjectures of possible errors in the energy scales based on comparison with other experimental results (``likely to be caused by errors in the energy scales''), as reported in the references cited by the authors of the Comment.
By the way, only one point at E$_{cm}$=2.63 MeV from the set of  \cite{Mazarakis73} enters the normalization with its 35$\%$ uncertainty, thus its weight is small. As for the direct measurement by \cite{Barron06}, these data have not been used in the normalization since they do not contribute in the E$_{cm}$= 2.5-2.63 MeV region. Thus, again the criticism raised by the authors of the Comment does not apply. Referring to the $\gamma$-ray measurements, the comparison with the $^{20}$Ne+$\alpha_{1}$ THM data has been done after correcting for the decay branching ratio. Data are all consistent within each other. Of course, new direct measurements are very welcome. The use of many data sets makes it possible to reduce the systematic error introduced by a single data set. It was recently proved in \cite{Lacognata2015} (where A.M.M. is coauthor) that having an extended normalization region and using more than one data set for normalization strongly reduce the influence of systematic errors affecting one data set.\\

3)We thank the authors to recall a fundamental principle of the quantum mechanics of what we are perfectly aware and we are sure our referees and the Editor are also. The normalization region is not affected by wrong J$^\pi$ assignment. The 2.567 MeV state is a J$^\pi$=0$^+$ (see Publisher Amendment). Thus, what is considered a major issue does not apply.
As for the odd spin states, whose assignment is taken from literature and done from visual inspection of excitation functions \cite{Abegg}, their spin is uncertain by +/-1. Previous studies cited in \cite{Abegg} give indeed tentative assignment of even neighboring values. Thus, their contribution in the modified R-matrix represents the average behavior of the two neighboring even values. 

 {}


\begin{thebibliography} {}
\bibitem{Akram}{Mukhamedzhanov A.M., Tang X. $\&$ D.Y. Pang, Comments on the $^{12}$C+$^{12}$C fusion S$^*$-factor, https://arxiv.org/pdf/1806.05921.pdf.}
\bibitem{Tumino2018}{Tumino A. {\textit et al.}, An increase in the $^{12}$C+$^{12}$C fusion rate from resonances at astrophysical energies. {\it Nature} \textbf{557}, 687 (2018).}
\bibitem{Akram1}{Mukhamedzhanov A.M., About $^{12}$C+$^{12}$C fusion astrophysical factor from Trojan horse method, https://arxiv.org/pdf/1806.08828.pdf.}
\bibitem{Spitaleri2011}{Spitaleri C. {\textit et al.}, The Trojan Horse Method in nuclear astrophysics. {\it Phys. At. Nucl.}, \textbf{74}, 1763 (2011).}
\bibitem{Tumino2013}{Tumino A. {\textit et al.}, New Advances in the Trojan Horse Method as an Indirect Approach to Nuclear Astrophysics. {\it Few Body Systems} \textbf{54}, 745 (2013).}
\bibitem{Tribble2014}{Tribble R. {\textit et al.}, Indirect techniques in nuclear astrophysics: a review. {\it Rep. Prog. Phys.} \textbf{77}, Issue: 10  106901 (2014).}
\bibitem{Spitaleri1999}{Spitaleri C. {\textit et al.}, Indirect $^7$Li(p,$\alpha$)$^4$He reaction at astrophysical energies. {\it Phys. Rev. C}, \textbf{60}, 055802 (1999).}
\bibitem{Lacognata2005}{La Cognata M. {\textit et al.}, Astrophysical S(E) factor of the $^{15}$N(p,$\alpha$)$^{12}$C reaction at sub-Coulomb energies via the
Trojan horse method. {\it Phys. Rev. C}, \textbf{76}, 065804 (2007).}
\bibitem{Lacognata2011}{La Cognata M. {\textit et al.}, The fluorine destruction in stars: first experimental study of the $^{19}$F(p,$\alpha_0$)$^{16}$O reaction at astrophysical energies. {\it Astrophysical J.}, \textbf{739}, L54 (2011). }
\bibitem{Lamia13}{Lamia L. {\textit et al.}, An updated $^{6}$Li(p,$\alpha$)$^{3}$He reaction rate at astrophysical energies with the Trojan horse method. {\it Astrophysical J.}, \textbf{768}, 65 (2013).} 
\bibitem{Tumino14}{Tumino A. \textit{et al.}, New determination of the $^2$H(d,p)$^3$H and $^2$H(d,n)$^3$He reaction rates at astrophysical energies. {\it Astrophysical J.}, \textbf{785}, 96 (2014).}
\bibitem{Pizzone17}{Pizzone R.G. {\textit et al.}, First measurement of the $^{19}$F($\alpha$,p)$^{22}$Ne reaction at energies of astrophysical relevance. {\it Astrophysical J.}, \textbf{836}, 57 (2017).}
\bibitem{Spitaleri17}{Spitaleri C. {\textit et al.}, Measurement of the $^{10}$B(p,$\alpha_{0}$)$^{7}$Be cross section from 5 keV to 1.5 MeV in a single experiment using the Trojan horse method. {\it Phys. Rev. C}, \textbf{95}, 035801 (2017).}
\bibitem{Dagata2018}{D'Agata G. {\textit et al.}, The $^{19}$F($\alpha$,p)$^{22}$Ne  reaction at energy of astrophysical relevance by means of the Trojan Horse Method and its implication in AGB stars.  {\it Astrophysical J.} \textbf{860}, 61-72 (2018). }
\bibitem{Dolinsky}{Dolinsky, E.I., Dzhamalov, P.O., $\&$ Mukhamedzhanov, A.M. Peripheral-model approach to stripping into resonant states, {\it Nuclear Physics A}, \textbf{202}, 97 (1973).}
\bibitem{Belyaeva01}{Belyaeva T.L. $\&$ Zelenskaya N.S., Quasimolecular states in $^{24}$Mg and d-$\alpha$ angular correlations in the $^{12}$C($^{14}$N,d)$^{24}$Mg$^*$($\alpha$)$^{20}$Ne reaction, {\it Phys. Rev. C}, \textbf{66}, 034604 (2002).}
\bibitem{Belyaeva02}{Belyaeva T.L., Zelenskaya N.S. $\&$ Aguero Granados M., Investigation of quasimolecular states in $^{24}$Mg$^{*}$Mg through the analysis of the angular d-$\alpha$ correlations in the $^{12}$C($^{14}$N,d)$^{24}$Mg($\alpha$)$^{20}$Ne reaction. {\it Phys. At. Nucl.}, \textbf{65}, 1657 (2002).}
\bibitem{Zurmuhle94}{Zurm\"uhle R.W.{\textit et al.}, Observation of $^{12}$C cluster transfer by angular correlation measurements. {\it Phys. Rev. C} \textbf{49}, 2549 (1994).}
\bibitem{Mazarakis73}{Mazarakis M.G. $\&$ Stephens W.E., Experimental measurements of the $^{12}$C+$^{12}$C nuclear reactions at low energies. {\it Phys. Rev. C}  \textbf{7}, 1280 (1973).}
\bibitem{High77}{High M.D., Cujec  B.,  The $^{12}$C+$^{12}$C sub-Coulomb fusion cross section. {\it Nucl. Phys. A}, \textbf{282}, 181 (1977).}
\bibitem{Kettner80}{Kettner K.U., Lorenz-Wirzba H., Rolfs C., The $^{12}$C+$^{12}$C reaction at subcoulomb energies. {\it Z. Phys. A}  \textbf{298}, 65 (1980).}
\bibitem{Barron06}{Barr\'on-Palos L. {\textit et al.}, Absolute cross sections measurement for the $^{12}$C+$^{12}$C system at astrophysically relevant energies. {\it Nucl. Phys. A} \textbf{779}, 318 (2006).}
\bibitem{Spillane07}{Spillane T. {\textit et al.}, $^{12}$C+$^{12}$C fusion reactions near the Gamow energy. {\it Phys. Rev. Lett.} \textbf{98}, 122501 (2007).}
\bibitem{CF88}{Caughlan G.R. $\&$ Fowler W.A., Thermonuclear reaction rates V. {\it At. Data Nucl. Data Tables}  \textbf{40}, (1988) 283.}
\bibitem{Lacognata2015}{La Cognata M. {\textit et al.}, Updated THM astrophysical factor of the $^{19}$F(p,$\alpha_0$)$^{16}$O reaction and influence of new direct data at astrophysical energies. {\it Astrophysical J.}, \textbf{805}, 128 (2015).} 
\bibitem{Abegg}{Abegg R. $\&$ Davis C.A., $^{24}$Mg states observed via $^{20}$Ne($\alpha$,$\alpha_{0}$)$^{20}$Ne. {\it Phys. Rev. C} \textbf{43}, 6 (1991).}
\bibitem{Pizzone2009} {Pizzone R.G. {\textit et al.}, Effects of distortion of the intercluster motion in $^2$H, $^3$He, $^3$H, $^6$Li, and $^9$Be on Trojan horse applications {\it Phys. Rev. C}, \textbf{80}, 025807 (2009).}
\bibitem{Guardo2017} {Pizzone R.G. {\textit et al.}, Assessing the near threshold cross section of the $^{17}$O(n,$\alpha$)$^{14}$C reaction by means of the Trojan horse method. {\it Phys. Rev. C}, \textbf{95}, 025807 (2017).}
\bibitem{Tumino2006} {Tumino A. {\textit et al.}, Validity test of the Trojan Horse Method applied to the $^7$Li+p$\rightarrow$$\alpha$+$\alpha$ reaction via the $^3$He break-up. {\it Eur. Phys. Jour. A}, \textbf{27}, 243 (2006).}
\bibitem{Pizzone2011} {Pizzone R.G. {\textit et al.}, Trojan horse particle invariance studied with the $^6$Li(d,$\alpha$)$^4$He and $^7$Li(p,$\alpha$)$^4$He reactions. {\it Phys. Rev. C}, \textbf{83}, 045801 (2011).}
\bibitem{Pizzone2013} {Pizzone R.G. {\textit et al.}, Updated evidence of the Trojan horse particle invariance for the $^2$H(d,p)$^3$H reaction. {\it Phys. Rev. C}, \textbf{87}, 025805 (2013).}

\end{thebibliography}
\end{document}